\documentclass{elsart}
\usepackage{amssymb}
\begin{document}
\begin{frontmatter}
\title{Dispersion relations in differential form\thanksref{talk}}
\thanks[talk]{Presented by Jan Fischer}
\author{Pavel Kol\'{a}\v{r} and Jan Fischer}
\address{Institute of Physics, Academy of Sciences of the Czech Republic, Na
Slovance 2, CZ-182 21  Praha 8, Czech Republic}
\begin{abstract}
Various forms of derivative dispersion relations, in which the dispersion
integral is replaced by a series of derivatives of the imaginary part of
a scattering amplitude, are reviewed. Conditions of their validity and
practical applicability as well as their relevance to high-energy
small-angle hadron-hadron scattering are discussed.
\end{abstract}
\end{frontmatter}
\thispagestyle{empty}

\section{Introduction}
There have been attempts since the mid 1970's to adapt the
existing dispersion relation technique to the energy range which
is far enough from resonance peaks and in which changes are slow
and cross sections smooth. Present experimental projects proposing
to measure small-angle high-energy hadron-hadron scattering at LHC
energy make this subject, after a certain time of silence, again
topical. Time therefore seems to be ripe to discuss the
high-energy status of dispersion relations, to point out a number
of remarkable merits of the differential approach, and also to
remind the reader of its limits and dangerous points which may
emerge at a careless application.

We cannot give here a comprehensive review of the subject. We will
just select several typical theorems to illustrate the variety of
results obtained in the past, referring for mathematical and
technical details to original papers. To discuss the issue, we
choose the example of $F(s,t)$, the crossing-even
amplitude of a generic hadron-hadron scattering process.

Let us consider a fixed-$t$ dispersion relation for $F(s,t)$,
\begin{equation}
 R(s) = \frac{2s^2}{\pi}\;\mathrm{P} \int_{s_0}^{\infty} \frac{\mathrm{d}s'}%
 {s'(s'^{2}-s^2)}\;I(s') \label{DR}
\end{equation}
(with poles and subtraction constants removed for simplicity),
where $R(s)$ and $I(s)$ is a shorthand for $\mathrm{Re}\,F(s,t)$
and $\mathrm{Im}\;F(s,t)$ respectively the $t$-dependence being
suppressed in the notation. It was proposed \cite{Bronzan} to
replace (\ref{DR}) by the ``quasilocal'' relation
\begin{equation}
\frac{R(s)}{s^\alpha} = \tan \left[\frac{\pi}{2} \left(\alpha - 1
+ \frac{\mathrm{d}}{\mathrm{d} \ln s} \right)\right]
\frac{I(s)}{s^{\alpha}} \label{DARal}
\end{equation}
with $\alpha$ real and $s$ the c.m. scattering energy squared. As
$s$ is linear in $E$, the laboratory scattering energy, $s$ and
$s'$ in (\ref{DR}) and (\ref{DARal}) can be replaced by $E$ and
$E'$ respectively, with inessential changes in the form of these
equations.

Choosing $\alpha = 1$ for simplicity, we obtain from (\ref{DARal})
\begin{equation}
g(x) = T(x) ,
\label{DAR1}
\end{equation}
where we use the notation $x=\ln s$, $g(x)=R(s)/s$, and $f(x)=I(s)/s$.
The ``tangent series'' $T(x)$ is
defined by the relation
\begin{eqnarray}
\label{notace}
T(x) &=& \tan \left(\frac{\pi}{2} \frac{\mathrm{d}}{\mathrm{d}x}
\right) f(x) =
 \sum_{n=1}^{\infty} a_{n} f^{(2n-1)}(x)  \\
\mbox{with} \,\,\,\,\,\, a_{n} &=&
\frac{2\pi^{2n-1}(2^{2n}-1)}{(2n)!} \,|B_{2n}|\, ,\nonumber
\end{eqnarray}
where $B_{2n}$ are the Bernoulli numbers.

The method of the derivative dispersion relations
(called also derivative analyticity relations) is again becoming
topical. It is therefore worth emphasizing their interesting
merits as well as categorical caveats, in particular

\begin{itemize}
\item the rather restrictive conditions of their validity
\item the problem of how to give (\ref{notace}) precise
mathematical meaning, and
\item problems of their practical applicability.
\end{itemize}

After a short historical survey, we shall discuss these subjects.
In discussing the relations (\ref{DARal}) to (\ref{notace}), we
have carefully to distinguish two very different formulations of
the problem: (i) either we keep the energy fixed and push the
approximation order (i.e. the number of terms in (\ref{notace}))
to infinity or (ii) the order is kept fixed and the energy tends
to infinity. Needless to say, the latter does not require so many
restrictive assumptions.

\section{History}
Applications of the ``derivative dispersion relation'' are very
wide. Already in 1968, Gribov and Migdal \cite{Gribov} made use of
this relation in the context of Regge theory. Later Bronzan, Kane
and Sukhatme \cite{Bronzan} introduced the method into the
phenomenology of high-energy small-angle hadron-hadron scattering.
In 1975 Kang and Nicolescu \cite{Nicol} proposed a model based on the
derivative relation to analyze the rising  total cross sections
for hadron-proton scattering.

Soon after \cite{Bronzan} was published it was shown that the
relation (\ref{DARal}) is restricted to certain mathematical
models; it was proved by Eichmann and Dronkers \cite{Eichmann}
that relation (\ref{DARal}) is exactly valid only on some class of
entire functions of $\ln s$. Bronzan, Kane and Sukhatme made the
crucial step towards applications, approximating (\ref{notace}),
the infinite series for $T(x)$, with a finite number of terms at a
fixed energy $s = \mathrm{e}^x$. As Eichmann and Dronkers
\cite{Eichmann} showed, however, the mathematical condition for the
convergence of the series excludes many cases of practical
interest.

In 1976 G. H\"{o}hler \cite{Hohler} and A. Bujak and O. Dumbrajs
\cite{Bujak} published critical comments on the use of the
derivative dispersion relations, showing that the difference
between a dispersion relation and its ``differential form''
(\ref{DARal}) may grow in an uncontrollable way, e.g. due to
low-energy contributions. In a series of papers P. Kol\'{a}\v{r},
J. Fischer and I. Vrko\v{c}
\cite{KolarPL,KolarPR,KolarJMP,Vrkoc,KolarCJP} gave the derivative
dispersion relations precise meaning and found conditions of their
validity and practical applicability.

In 1986 J.C. Pumplin, W.W. Repko, G.L. Kane and M.J. Duncan
\cite{Duncan} used the method to study the gluonic production of
vector bosons and boson pairs in the Standard Model.

In 1990 and 1999 M.N. Mnatsakanova and Yu.S. Vernov \cite{Vernov}
applied the method to weakly oscillating amplitudes and
established validity conditions of the derivative relations. M.J.
Menon and co-authors \cite{Menon} used the method for the case of
an arbitrary number of subtractions, and made a systematic
comparison of derivative relations with experimental data.

\section{Convergence of the tangent series $T(x)$ at a fixed, finite energy}
{\bf Theorem 1} \cite{KolarJMP}: Let $f$: $\Rset^{1} \rightarrow \Rset^{1}$. The series
$T(x)$, (\ref{notace}), is convergent at a point $x \in \Rset^{1}$ if and only if
the series
\begin{equation}
\label{sumder}
 D(x) = \sum_{n=0}^{\infty} f^{(2n+1)}(x)
\end{equation}
is convergent.

This relatively simple theorem is of fundamental importance for
many subsequent results. As for practical applications, we have to
parameterize a scattering amplitude in an energy interval. So we
need the following

{\bf Theorem 2} \cite{Vrkoc}: Let $f$: $I \rightarrow \Rset^{1}$
have all derivatives at every $x \in I$, $I \subset \Rset^{1}$
(i.e., let $f \in C^{\infty}(I)$). If $T(x)$ converges for every
$x \in I \subset \Rset^{1}$, then an entire function of complex
$x$ exists which assumes the values of $f(x)$ on $I$.

These results show an extraordinarily restricted validity of the
``derivative dispersion relations'' at finite energy: it follows
that $T(x)$ is convergent on an energy interval $I \subset
\Rset^{1}$ only if $f(x)$ is an entire function of complex  $x$
and if two series, (\ref{sumder}) for $D(x)$ and
$E(x)=\sum_{n=0}^{\infty} f^{(2n)}(x)$, also converge.

\section{Link to dispersion relations}
If $T(x)$ is convergent, we can derive the corresponding dispersion
relation:
\begin{equation}
\label{linkT} T(x) = \tan \left(\frac{\pi}{2}
\frac{\mathrm{d}}{\mathrm{d}x} \right) f(x) =
\int_{0}^{\infty}a(t) \e^{-t} \left[f(x+t)-f(x-t)
\right]\:\mathrm{d}t
\end{equation}
with
\begin{equation}
a(t) = \frac{2}{\pi} \left(1-\e^{-2t}\right)^{-1}
\label{linka}
\end{equation}
and
\begin{equation}
D(x) = \sum_{n=0}^{\infty}f^{(2n+1)}(x) =
\frac{1}{2}\int_{0}^{\infty} \e^{-t} \left[f(x+t)-f(x-t)
\right]\;\mathrm{d}t . \label{linkD}
\end{equation}
Setting $x = \ln s$ in (\ref{linkT}), we obtain
\begin{equation}
\tan \left(\frac{\pi}{2} \frac{\mathrm{d}}{\mathrm{d} \ln s}
\right) f(\ln s) = \frac{2s}{\pi}\int_{0}^{\infty}\frac{f(\ln
s')}{s'^{2}-s^{2}}\;\mathrm{d}s' . \label{linklogs}
\end{equation}
This is to be confronted with the ordinary dispersion relation,
which is obtained by putting $f(x) = \mathrm{Im}\,F(s)/s$.

\section{Practical applicability of the derivative dispersion relations}

The results discussed in the previous sections indicate that the class of
``amplitudes'' to which the derivative relations may be applied is very narrow.
Problems of their practical applicability were studied in detail in
\cite{Hohler,Bujak,KolarJMP}, and we shall briefly discuss some of the results
obtained.

Let us consider two fits of the imaginary part, $\mathrm{Im}\,F^{D}$
and $\mathrm{Im}\,F^{B}$, which are used in the dispersion relation
approach and in the derivative approach respectively. Can one
impose an upper bound on the modulus of their difference? (Note
that $\mathrm{Im}\,F^{D}$ belongs to a much wider class of functions
than $\mathrm{Im}\,F^{B}$.)

It is pointed out in \cite{Hohler,Bujak,KolarJMP} that
$\mathrm{Im}\,F^{D}-\mathrm{Im}\,F^{B}$ may, in certain situations,
grow in an uncontrollable way due to low-energy contributions.
Bounds can be obtained only if bounds on low-energy contributions
are known; see the cited papers for details. A simple example to
illustrate the situation is as follows \cite{Bujak}. Let us add
the term  $c s^{\alpha}$ to a parametrization of the imaginary
part of the scattering amplitude, with $0<\alpha \leq 1$ and $|c|$
very small, so as not to change the fit. Then the real part
acquires the term $-c s^{\alpha} \cot(\alpha \pi /2)$, which
becomes arbitrarily large for $\alpha$ near zero. This is an
argument against the derivative relations, but not only against
them, because similar problems can arise also in an ordinary
dispersion relation \cite{KolarJMP}.

We emphasize once again that the class of functions for which the derivative
dispersion relations are applicable is very narrow; moreover, predictions
based on them may be unstable.

A dramatic change takes place when we pass from a finite energy
relation to the high-energy limit. A wide spectrum of derivative
relations are valid in the high-energy limit for a large class of
functions. More than that, the infinite series (\ref{notace}) for
$T(x)$ can be safely replaced by its first term!

\section{Derivative dispersion relations in the high-energy limit}
The validity of such relations has been proved
\cite{KolarPL,KolarPR} for a class of functions which is almost as
large as the class of scattering amplitudes described by ``first
principles'', i.e., functions satisfying analyticity, crossing
symmetry, polynomial boundedness, etc. Then it is sufficient to
retain the first term for $T(x)$, $3\pi |B_{2}| f'(x)$, and take
the limit $s \rightarrow \infty$. In this way, a number of
high-energy ($s \rightarrow \infty$) derivative relations are
obtained. Some of them correlate the real with the imaginary part of $F(s)$,

\begin{equation}
\frac{\mathrm{Re}\,F(s)}{s} \,\,\, \rightarrow \,\,\,
\frac{\pi}{2}\;\frac{\mathrm{d}}{\mathrm{d} \ln s}\,
\frac{\mathrm{Im}\,F(s)}{s} \,\,\, , \label{ReIm}
\end{equation}
others the phase with the modulus of the amplitude,
\begin{equation}
\frac{\mathrm{d}}{\mathrm{d} \ln s} \,\, \ln
\left|\frac{F(s)}{s}\right| \,\, \rightarrow \,\,
\frac{2}{\pi}\;\arctan\left( \frac{\mathrm{Re}\,
F(s)}{\mathrm{Im}\,F(s)}\right) \,\,\, . \label{phasemod}
\end{equation}
The arrow $\rightarrow$ means either that the ratio of the left to the
right-hand side tends to unity, or that their difference tends to zero with
$s \rightarrow \infty$.

Let us give one example to illustrate the results.

{\bf Theorem 3} \cite{KolarPR}: Let $f(s)$ satisfy, apart from the
properties of analyticity, crossing symmetry, polynomial
boundedness, positivity of the imaginary part and the
Froissart-Martin bound, the following conditions:

\begin{equation}
\int_{s_1}^{\infty}\;\mathrm{Im}F(s)\;\frac{\mathrm{d}s}{s}\;=\;\infty
\end{equation}
for some $s_1>0$ and
\begin{equation}
\lim_{s \rightarrow \infty} \left| \ln \left|\frac{F(s)}{s^{1-a}}\right| \right| = \infty
\label{limln}
\end{equation}
for some real $a$. If the limit $\lim_{s \rightarrow \infty}A(s)$ exists, where
\begin{equation}
A(s) = \frac{\mathrm{d}}{\mathrm{d}\ln s} \ln
\left|\frac{F(s)}{s^{1-a}} \right| \left/
\left[a+\frac{2}{\pi}\arctan\left(\frac{\mathrm{Re}\,F(s)}{\mathrm{Im}\,F(s)}\right)
\right]\right.,
\end{equation}
then it is equal to 1.

Taking $a=0$, we obtain
\begin{equation}
\frac{\mathrm{d}}{\mathrm{d} \ln s} \ln \left|\frac{F(s)}{s}
\right| \left/
\arctan\left(\frac{\mathrm{Re}\,F(s)}{\mathrm{Im}\,F(s)}\right)\right.\,\,
\mathop{\longrightarrow}_{s \rightarrow \infty} \,\,\,
\frac{\pi}{2}\,\,\, .
\end{equation}

As was mentioned above, there are a number of analogous asymptotic relations
connecting the real with the imaginary part, phase with modulus, both for the
crossing-even and the crossing-odd scattering amplitude. Details can be found
in \cite{KolarPR} and \cite{KolarCJP}.

\section{Conclusions}

1. The derivative dispersion relations again become topical, even
more than ever before, because of the extraordinarily high energy
on the LHC.

2. In using derivative dispersion relations, there are essentially
two different approaches. One is to keep the energy fixed and to
approximate the tangent series $T(x)$ with a finite, possibly
increasing, number of terms. The other is to keep fixed the number
of terms approximating the series $T(x)$ and let the scattering
energy tend to infinity. The class of applicability in the
latter approach is considerably wider than that in the former.

3. If the tangent series $T(x)$ converges on an interval $I$,
$f(x)$ must be extensible to an entire function of $x = \ln s$.
Then the sum of $T(x)$ is equal to the corresponding dispersion
integral. \emph{Conclusion}: This class of functions is too narrow
to contain the true amplitude.

4. If $f \in C^{\infty}$ and the dispersion integral converges,
the latter is equal to the generalized sum (\ref{linkT}) (modified
Borel summation). This class is wider, but the true scattering
amplitude is not necessarily included.

5. In the high energy limit, $s \rightarrow \infty$, derivative
dispersion relations are valid in which the ``tangent'' operator
$T(x)$ is replaced by its first expansion term. The class of
applicability includes the majority of physically interesting
functions.

We thank Vojt\v{e}ch Kundr\'{a}t for organizing this nice and
successful Blois conference 2001.


\begin{thebibliography}{00}
\bibitem {Bronzan} J.B. Bronzan, Argonne Nat.Lab. Report No ANL/HEP 7327, 1973;\\
J.B. Bronzan, G.L. Kane and U.P. Sukhatme, Phys.Lett. \textbf{49B}
(1974) 272.
\bibitem{Gribov} V.N. Gribov and A.B. Migdal, Yad.Fiz. \textbf{8} (1968) 1213;\\
Sov.J.Nucl.Phys. {\bf 8}(1969) 702.
\bibitem{Nicol} Kyungsik Kang and B. Nicolescu, Phys.Rev \textbf{D 11} (1975) 2461.
\bibitem{Eichmann} G.K. Eichmann and J. Dronkers, Phys.Lett. \textbf{B 52} (1974)
428.
\bibitem{Hohler} G. H\"{o}hler, {\it in} Landolt-B\"{o}rnstein, New Series,
Group I, Vol. 9b, Part 2, edited by H. Schopper (Springer, Berlin, 1983), p. 41.
\bibitem{Bujak} A. Bujak and O.Dumbrajs, J.Phys. \textbf{G 2} (1976) L 129.
\bibitem{KolarPL} J. Fischer and P. Kol\'{a}\v{r}, Phys.Lett. \textbf{64B} (1976)
45.
\bibitem{KolarPR} J. Fischer and P. Kol\'{a}\v{r}, Phys.Rev. \textbf{D17} (1978)
2168.
\bibitem{KolarJMP} P. Kol\'{a}\v{r} and J. Fischer, J.Math.Phys. \textbf{25}
(1984) 2538.
\bibitem{Vrkoc} I. Vrko\v{c}, Czech. Math. J. \textbf{35} (1985) 59.
\bibitem{KolarCJP} J. Fischer and P. Kol\'{a}\v{r}, Czech.J.Phys. \textbf{B37} (1987) 297.
\bibitem{Duncan} M.J. Duncan and G.L. Kane, Nucl.Phys \textbf{B 272} (1986) 517.
\bibitem{Vernov}  Yu.S. Vernov and M.N. Mnatsakanova, in
\textit{Proc. of the XII Intern. Seminar on high energy physics
and quantum field theory, Protvino}, (Nauka 1990)\\
Yu.S. Vernov and M.N. Mnatsakanova, in \textit{Proc. of the VIII
Blois Workshop, Protvino, Russia}, (World Scientific, Singapore,
2000).
\bibitem{Menon} M.J. Menon, A.E. Motter and B.M. Pimentel, hep-th/9810196 v2,\\
A.F. Martini, M.J. Menon, J.T.S. Paes and M.J. Silva Nieto, hep-ph/9810498 v2,\\
R.F.\ \'{A}vila, E.G.S.\ Luna and M.J.\ Menon, hep-ph/0105065.
\end{thebibliography}
\end{document}